\documentclass[aps,prx,preprint,showpacs,preprintnumbers,amsmath,amssymb,floatfix,longbibliography]{revtex4-2} 

\usepackage[dvipdfmx]{graphicx}
\usepackage{bm}
\usepackage{xcolor}

\newcommand{\be}{\begin{equation}}
\newcommand{\ee}{\end{equation}}
\newcommand{\eq}[1]{Eq.~(\ref{#1})}

\def\bea{\begin{eqnarray}}
\def\eea{\end{eqnarray}}

\def\bra{\langle}
\def\ket{\rangle}
\def\vq{{\bf q}}

\def\vk{{\bf k}}

\def\vr{{\bf r}}
\def\qp{{\bf q}_{\parallel}}

\begin{document}

\title{Low-energy plasmon excitations in infinite-layer nickelates} 

\author{ Luciano Zinni$^{1}$, Mat\'{\i}as Bejas$^{2}$, Hiroyuki Yamase$^{3}$, and Andr\'es Greco$^{2}$}
\affiliation{
{$^1$}Facultad de Ciencias Exactas, Ingenier\'{\i}a y Agrimensura,
Avenida Pellegrini 250, 2000 Rosario, Argentina\\
{$^2$}Facultad de Ciencias Exactas, Ingenier\'{\i}a y Agrimensura and
Instituto de F\'{\i}sica Rosario (UNR-CONICET),
Avenida Pellegrini 250, 2000 Rosario, Argentina\\
{$^3$}International Center of Materials Nanoarchitectonics, 
National Institute for Materials Science, Tsukuba 305-0047, Japan\\
}

\date{\today}

\begin{abstract}
The discovery of superconductivity in infinite-layer nickelates is presently an important topic in condensed-matter physics, and potential similarities to and differences from cuprates are under intense debate. 
We determine general features of the charge excitation spectrum in nickelates from two opposite viewpoints:
(i) Nickelates are regarded as strongly correlated electron systems like cuprate superconductors and thus can be described by the $t$-$J$ model, and
(ii) electron correlation effects are not as strong as in cuprates, and thus, random-phase approximation (RPA) calculations may capture the essential physics.
We find that in both cases, plasmon excitations are realized around the momentum transfer $\vq=(0,0,q_z)$, although they tend to be damped more strongly in the RPA.
In particular, this damping is enhanced by the relatively large interlayer hopping expected in nickelates.
Besides reproducing the  optical plasmon at $\vq=(0,0,0)$ observed in Nd$_{0.8}$Sr$_{0.2}$NiO$_2$, we obtain low-energy plasmons with gaps of $\sim 360$ and $\sim 560$ meV at $\vq=(0,0,q_z)$ for finite $q_z$ in cases (i) and (ii), respectively.  
The present work offers a possible theoretical hint to answer whether nickelates are cupratelike or not and contributes to the general understanding of the charge dynamics in nickelates.
\end{abstract}

\maketitle
\section{Introduction}

The study of unconventional superconductivity in strongly correlated systems is one of the major topics in solid-state physics.
Thirty-five years after their discovery\cite{bednorz86}, cuprates still attract a lot of attention, not only for the high value of their superconducting transition temperature $T_c$ but also for their many surprising normal-state properties \cite{keimer15,timusk89}.
The interest in understanding the physics of cuprates and unconventional superconductivity also boosted the study of other materials such as cobaltates \cite{takada03}, organic superconductors \cite{mcKenzie97}, Fe pnictides as well as heavy fermions \cite{scalapino12}, and recently discovered kagome materials \cite{ortiz19}. 

Remarkably, a new discovery has occurred in the last few years.
We have entered in what some people call the ``nickel age of superconductivity'' \cite{norman20,pickett21}.
Nickel is next to Cu in the periodic table, and both have $3d$ active orbitals, suggesting possible similarities to cuprates.
Along these lines, it was already argued \cite{anisimov99} in 1999 that infinite-layer nickelates could be analogous to cuprates.
However, only recently was superconductivity with $T_c= 9$-$15$ K reported in thin films of the hole-doped infinite-layer nickelate Nd$_{0.8}$Sr$_{0.2}$NiO$_2$ \cite{li19}.
The occurrence of superconductivity in infinite-layer nickelates is now also confirmed for hole-doped PrNiO$_2$ \cite{osada20} and LaNiO$_2$ \cite{osada21}.

In spite of possible analogies between nickelates and cuprates the discussion about the similarities and differences between them is substantial \cite{pascut22}.
For instance, it was suggested that doped holes in nickelates go mainly to the Ni $3d_{x^2-y^2}$ orbital \cite{goodge21,rossi21}, while it is known that doped holes in cuprates go to the O $2p$ orbitals.
In this aspect, hole-doped nickelates might be closer to electron-doped cuprates, where electrons reside mostly at Cu \cite{ament11}.

One important difference from cuprates is the fact that undoped nickelates with a half-filled $3d_{x^2-y^2}$ band do not exhibit antiferromagnetic long-range order \cite{ortiz22}; instead, they are found to be paramagnetic and weakly metallic \cite{li19}.
Nevertheless, antiferromagnetic fluctuations were observed by NMR \cite{cui21,zhao21}.
In addition, resonant inelastic x-ray scattering (RIXS) experiments at the Ni $L_3$ edge have detected paramagnons for several dopings which are similar to those observed in cuprates \cite{letacon11}, with substantial nearest-neighbor coupling $J \sim 64$ meV \cite{lu21,hepting21b,katukuri20}.
Also, as in cuprates, nickelates show a $T_c$ dome as a function of doping \cite{li20}, and the normal-state resistivity behaves similarly to cuprates for the whole phase diagram \cite{Klee22}.

Furthermore, infinite-layer nickelates $R$NiO$_2$ (with the rare-earth ions $R$ $=$ La, Pr, Nd) are isostructural to the infinite-layer cuprate CaCuO$_2$, both having missing apical oxygen, making the systems rather two-dimensional \cite{botana20}.
Moreover, both contain planes of NiO$_2$ (CuO$_2$) separated by planes of $R$ (Ca) ions.
For the case of CaCuO$_2$ the low-energy physics can be described by a one-band correlated model, for instance, the $t$-$J$ model.
The case of $R$NiO$_2$ is much more controversial \cite{chen22,karp21}.
While some authors propose a one-band model description \cite{held22,kitatani20,karp20} for infinite-layer nickelates, others suggest a multiorbital picture, where in addition to Ni $d_{x^2-y^2}$ and O $p_\sigma$, other Ni $d$ \cite{kreisel22,lechermann20} and Ni $s$ \cite{plienbumrung22,peng21} orbitals should be included.
In addition, $R$ orbitals may play a direct role \cite{lechermann20}.
An effective two-band model supported by x-ray spectroscopy on the parent compound indicates that the NiO$_2$ planes host strongly correlated Ni $d_{x^2-y^2}$ states, but the $R$ interlayers give rise to a three-dimensional weakly interacting $5d$ metallic state \cite{hepting20}.
It is also thought that the $R$ orbital acts as a self-doping band \cite{held22}.
The role of correlations in the Ni $d$ bands is discussed in Ref. \cite{kreisel22}, where it is argued that while the Ni $d_{x^2-y^2}$ band is correlated, the Ni $d_{z^2}$ band has itinerant character, along with Ni $s$ \cite{plienbumrung22}.
As a consequence, nickelates might be not cupratelike.
In contrast, NdNiO$_2$ and CaCuO$_2$ were studied by a cluster dynamical mean field theory (DMFT) \cite{karp22}, and the authors arrived at the conclusion that nickelates are cupratelike systems, i.e., that nickelates can be considered strongly correlated systems.

Cuprates and nickelates are layered systems.
Thus, besides the usual optical plasmons, acoustic plasmons are expected, as theoretical models suggest.
Acoustic plasmons were studied in weak-coupling theory a long time ago \cite{grecu73,fetter74,grecu75}.
Plasmons have a momentum transfer $\vq=(q_x,q_y,q_z)$, where $(q_x,q_y)$ defines the in-plane momentum $\vq_{\parallel}$ and $q_z$ is the out-of-plane momentum.
A plasmon branch is specified by $q_z$ in the layered system.
For $q_z=0$ the optical plasmon branch is realized, which can be probed in optical experiments at $\vq=(0,0,0)$.
For finite $q_z$ the acoustic plasmon branches show a fast decrease in energy at low $\vq_{\parallel}$.
Cuprates are layered and correlated systems, and recent advances in high-resolution  RIXS experiments allow to detect low-energy-plasmon excitations in both electron- and hole-doped cuprates \cite{hepting18,jlin20,nag20,hepting22,singh22}.
These plasmons were well described in the framework of the layered $t$-$J$ model with long-range Coulomb interaction \cite{greco16,greco19,greco20,nag20,hepting22}, namely, the $t$-$J$-$V$ model (see Appendix A).
Notably, it was revealed recently \cite{greco16,hepting22} that the low-energy plasmons in cuprates are not strictly acoustic, but have finite energy at $\vq_{\parallel}=(0,0)$, which was predicted to increase with increasing the interlayer hopping $t_z$ \cite{greco16}.  

The plasmon gap and $t_z$ are particularly large in the electron-doped infinite-layer cuprate Sr$_{0.9}$La$_{0.1}$CuO$_2$ (SLCO) \cite{hepting22}, which is isostructural to infinite-layer nickelates \cite{botana20,karp22}.
The observed RIXS plasmons in SLCO were accurately captured by calculations in the layered $t$-$J$-$V$ model \cite{hepting22} using the tight-binding fit to the Cu $d_{x^2-y^2}$ band at the Fermi surface reported in Ref.\cite{botana20} for CaCuO$_2$.
In Ref.\cite{botana20} a similar tight-binding fit to the Ni $d_{x^2-y^2}$ band for LaNiO$_2$ was also provided.
Then, if $R$NiO$_2$ and SLCO are isostructural and both are cupratelike, the $t$-$J$-$V$ model should yield low-energy plasmons for $R$NiO$_2$.
Importantly, an optical plasma frequency of about $1.05$ eV was measured very recently in Nd$_{0.8}$Sr$_{0.2}$NiO$_2$ \cite{cervasio22}.
This benchmark for the optical plasmon branch represents an important advantage because it is an indicator of the suitable strength of the long-range Coulomb repulsion in the material class (see Appendix A), thus allowing us to perform reliable $t$-$J$-$V$ model calculations for the low-energy plasmons in nickelates.

In this paper, in the framework of a one-band model, we study the emergence of low-energy plasmons in nickelates assuming two distinct cases:
(i) Nickelates are strongly correlated systems and,
(ii) nickelates are characterized by weak electron interactions.
The consideration of these two opposite cases and the obtained presence and absence of plasmons can help to identify whether nickelates fall into regime (i) or (ii) or a mixed regime. 
Moreover, the obtained plasmon dispersion relations and plasmon gap contribute to the understanding of the charge dynamics in nickelates, which is an integral part of these materials.

In Sec. II we show the results of our $t$-$J$-$V$ and random-phase approximation (RPA) calculations, and in Sec. III we present the discussion and conclusion.
In Appendix A we summarize the main formulas for the layered $t$-$J$-$V$ model which were extensively discussed in previous papers \cite{foussats04,bejas12,greco16}, and in Appendix B we provide the main expressions for the standard RPA calculation \cite{mahan}.

\maketitle

\maketitle
\section{Results}

In the following we first assume that nickelates fall into the strongly correlated limit\cite{karp22}, which corresponds to the case of  cuprates.
As a consequence, we apply the layered $t$-$J$-$V$ model (see Appendix A) to $R$NiO$_2$.
We use the tight-binding parameters proposed in Ref.\cite{botana20} for LaNiO$_2$, i.e., $t'/t=-0.25$, $t_z/t=0.084$.
It is important to remark that our choice of band parameters from Ref.[\onlinecite{botana20}] is motivated by the fact that the parameters provided in this paper for CaCuO$_2$ are very reliable for discussing the plasmons in SLCO\cite{hepting22}, especially the value for $t_z$ which is of great interest for the present study.
The value for $J$ is unknown in nickelates and it takes scattered values in the literature\cite{nomura22}; here we chose $J/t=0.3$. However, the value for $J$ plays a minor role when discussing plasmons \cite{greco16,bejas17} (see also Appendix A).
For the Coulomb potential we have two independent parameters, $V_c$ and $\alpha$.
Since the optical plasma frequency depends on $V_c/\alpha$ and $\alpha$ has no impact on the calculation of the low-energy plasmons \cite{hepting22}, we choose its value $\alpha = 3$ to be of the order of the value used in Refs.\cite{hepting22} and select the value of $V_c/t = 32$ that gives the observed $\omega_{op} \sim 1.05$ eV.
To convert the energy to eV we use $t/2=0.368$ eV \cite{botana20}, where the factor $1/2$ is required by the large-$N$ formalism \cite{greco16}.
In addition we use the broadening $\Gamma/t=0.1$ (see Appendix A), which is typically of the order of the broadening observed in RIXS experiments \cite{nag20,hepting22,jlin20}.  
The calculation is performed for hole doping $\delta=0.2$, and the temperature is fixed to $T=0$. 

\begin{figure}[ht]
\centering
\includegraphics[width=16cm]{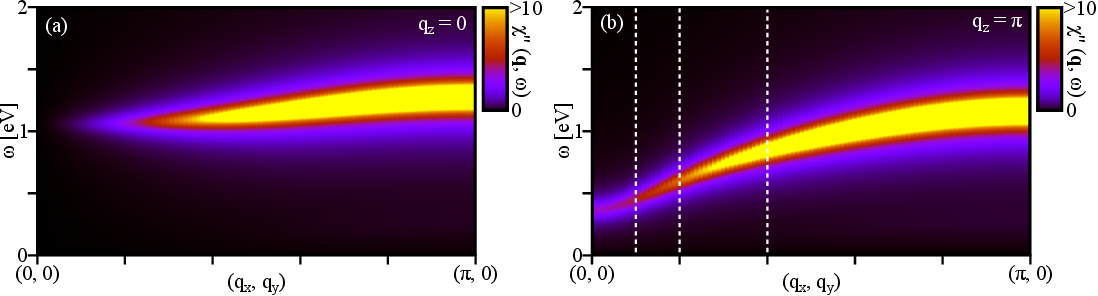}
\caption{(Color online) 
(a) and (b) Intensity color map for $\chi''_c({\bf q},\omega)$ for momenta along the $(0,0)$-$(\pi,0)$ direction for $q_z=0$ and $q_z=\pi$,  respectively.
The thin dotted lines in (b) mark the three momenta chosen in Fig. \ref{fig2}.
}
\label{fig1}
\end{figure}

In order to show the optical plasmon branch ($q_z=0$), in Fig. \ref{fig1}(a) we show a color intensity map of the charge-charge correlation function $\chi''_{c}(\vq,\omega)$ (see Appendix  A) in the plane $\vq_\parallel$-$\omega$ in the direction $(0,0)$-$(\pi,0)$.
The optical plasmon branch is rather flat and well defined for all $\vq_\parallel$.
The optical plasma frequency at $\vq=(0,0,0)$ takes the value $\omega_{op}\sim 1.05$ eV \cite{cervasio22}.
Figure \ref{fig1}(b) shows a similar intensity map for $q_z=\pi$, which yields plasmons with the lowest energy \cite{hepting18,greco19}. Like in Fig. \ref{fig1}(a), the plasmons are well defined for all $\vq_\parallel$.
Consistent with the large value of the interlayer hopping in the system ($t_z/t=0.084$) there is a plasmon gap of about $366$ meV at $(0,0,\pi)$, which is much larger than the observed plasmon gap of about $120$ meV in the cuprate SLCO \cite{hepting22}. 

\begin{figure}[ht]
\centering
\includegraphics[width=16cm]{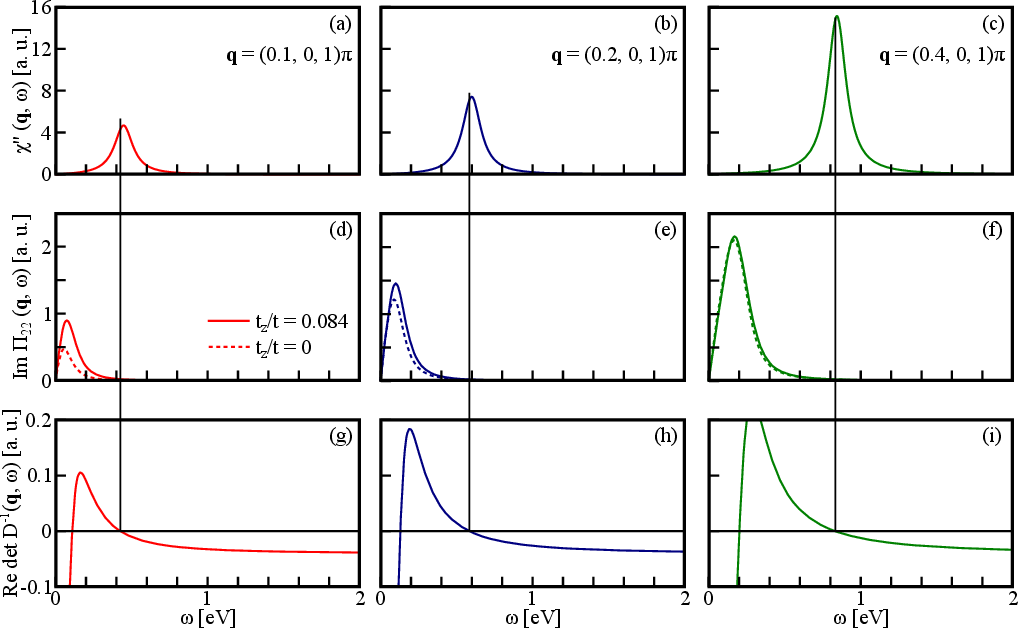}
\caption{(Color online) 
(a)-(c) $\chi''_c({\bf q},\omega)$ for momenta $(0.1, 0, 1)\pi$, $(0.2, 0, 1)\pi$, and $(0.4,0,1)\pi$, respectively.
(d)-(f) The particle-hole continuum for the corresponding momenta.
Dashed lines show the particle-hole continuum for $t_z/t=0$ for comparison with the case for $t_z/t=0.084$. 
(g)-(i) The determinant of the matrix $D^{-1}$ for the corresponding momenta. 
}
\label{fig2}
\end{figure}
\begin{figure}[ht]
\centering
\includegraphics[width=8cm]{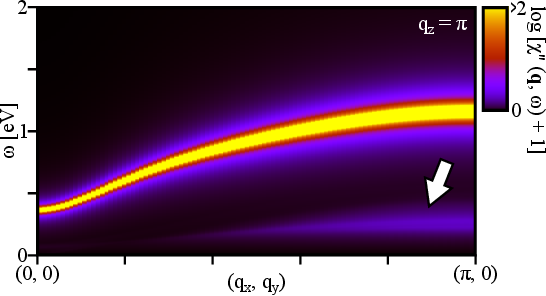}
\caption{(Color online) 
The same color map as in Fig. \ref{fig1}(b), but with a logarithmic intensity scale to emphasize the very weak particle-hole continuum, indicated by the white arrow, below the plasmon branch. 
}
\label{fig3}
\end{figure}

In Figs. \ref{fig2}(a)-\ref{fig2}(c) we show results for $\chi''_{c}(\vq,\omega)$ for the momenta $(0.1,0,1)\pi$, $(0.2,0,1)\pi$ and $(0.4,0,1)\pi$, respectively, indicated by thin dotted lines in Fig. \ref{fig1}(b).
The plasmons are well defined, in spite of the broadening $\Gamma/t=0.1$.
As expected, the intensity of the plasmon peak increases with increasing $\vq_\parallel$.
We note that there is no indication of a large particle-hole continuum below the plasmons in Fig. \ref{fig1}(a), \ref{fig1}(b) and Fig. \ref{fig2}(a)-\ref{fig2}(c).
Nevertheless, Fig. \ref{fig3} shows the intensity color map for $\chi''_c({\bf q},\omega)$ in a logarithmic scale for momenta along the $(0,0)$-$(\pi,0)$ direction and $q_z=\pi$, revealing the weak continuum at low energy, well separated from the plasmon branch.
Thus, a large charge spectral weight is carried dominantly by the collective plasmon excitations.
This minor role of the particle-hole continuum in the $t$-$J$-$V$ calculations will be discussed again later.

\begin{figure}[ht]
\centering
\includegraphics[width=16cm]{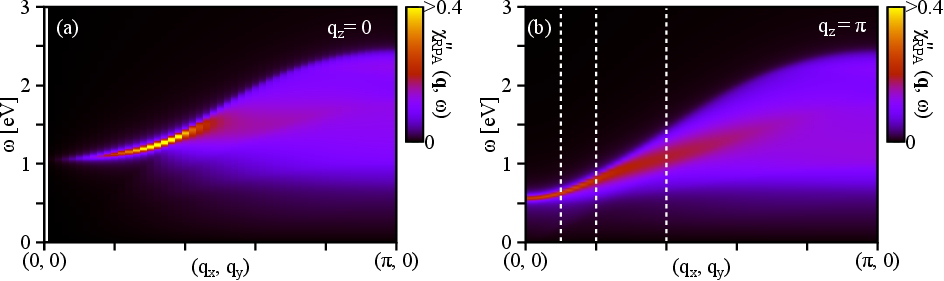}
\caption{(Color online) 
(a) and (b) Intensity color map of $\chi''_{RPA}({\bf q},\omega)$ for momenta along the $(0,0)$-$(\pi,0)$ direction at $q_z=0$ and $q_z=\pi$, respectively, computed in RPA.
The thin dotted lines in (b) mark the three momenta chosen in Fig. \ref{fig5}.}
\label{fig4}
\end{figure}

Next we discuss the case of weakly interacting electrons in the RPA method \cite{mahan} (see Appendix B).
We use the same parameters as those for the layered $t$-$J$-$V$ model \cite{botana20}, except for $V_c/t=16.5$ to adjust the optical plasma frequency at $1.05$ eV in the context of the RPA method.
To convert energies to eV we use $t=0.368$ eV \cite{botana20}. 

In Fig. \ref{fig4}(a) we show the intensity map for $\chi_{RPA} ''(\vq,\omega)$ in the plane $\vq_\parallel$-$\omega$ in the $(0,0)$-$(\pi,0)$ direction for the optical branch ($q_z=0$).
Like in Fig. \ref{fig1}(a), the optical plasma frequency at $\vq=(0,0,0)$ reproduces the observed experimental value \cite{cervasio22}.
The plasmon is well defined at low $\vq$, but in contrast to the $t$-$J$-$V$ model, the plasmon loses intensity beyond $\vq \sim (0.4,0,0)\pi$, when the plasmon branch mixes with the particle-hole continuum. 
Figure \ref{fig4}(b) shows the intensity map for $\chi_{RPA} ''(\vq,\omega)$ for $q_z=\pi$. 
The plasmons in RPA become less and less defined with increasing $\vq_\parallel$ because they are strongly affected by the particle-hole continuum and mix with the continuum even at low $\vq_\parallel$.  
The calculated plasmon gap in RPA is about $567$ meV, which is larger than that in the layered $t$-$J$-$V$ model calculation.

\begin{figure}[ht]
\centering
\includegraphics[width=16cm]{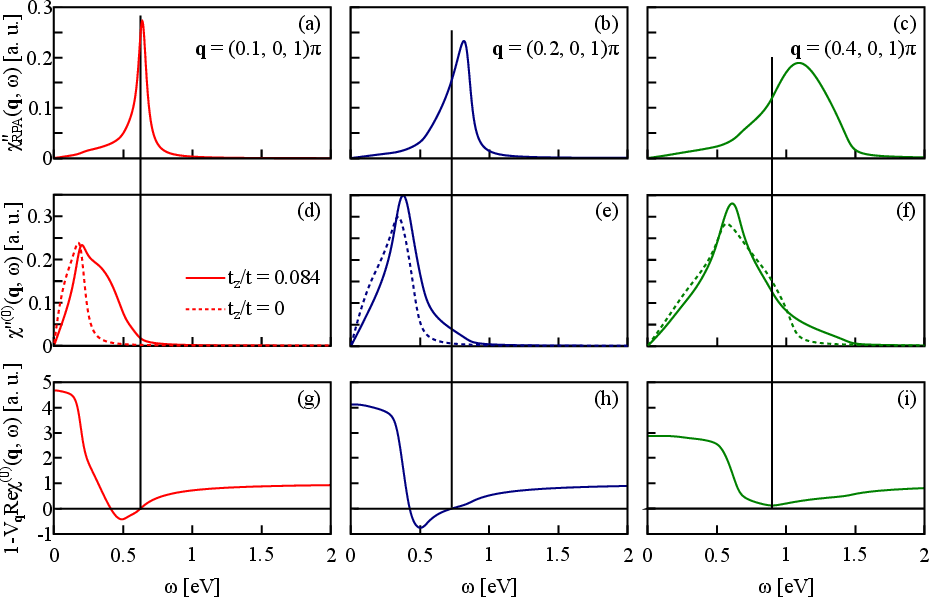}
\caption{(Color online) 
(a)-(c) $\chi''_{RPA} ({\bf q},\omega)$ for momenta $(0.1, 0, 1)\pi$, $(0.2,0,1)\pi$, and $(0.4,0,1)\pi$, respectively, calculated in the RPA.
(d)-(f) $\chi''^{(0)}({\bf q},\omega)$ for the same momenta as in (a)-(c).
Dashed lines show the particle-hole continuum for each momentum for $t_z/t=0$.
(g)-(i) show the denominator of the RPA charge susceptibility for each momentum.  
}
\label{fig5}
\end{figure}

In Figs. \ref{fig5}(a)-\ref{fig5}(c) we show $\chi_{RPA}''(\vq,\omega)$ for the momenta $(0.1,0,1)\pi$, $(0.2,0,1)\pi$, and $(0.4,0,1)\pi$, respectively.
The plasmons are not sharp and become broad with increasing $\vq_\parallel$.
In contrast to the results for the $t$-$J$-$V$ model [see Figs. \ref{fig2}(a)-\ref{fig2}(c)], the intensity of the maximum of
$\chi''_{RPA}(\vq,\omega)$ decreases with increasing $\vq_\parallel$.
In Figs. \ref{fig5}(d)-\ref{fig5}(f) we show the particle-hole continuum $\chi''^{(0)}(\vq,\omega)$ for each $\vq$.
For comparison we had added in each plot the corresponding particle-hole continuum for the same conditions but for $t_z/t=0$ (dashed lines).
The extended continuum for large $t_z$ affects the low-energy plasmon formation at low $\vq_\parallel$.
Consequently, for $q_z=\pi$, as well as for any finite $q_z$ (not shown), there is a large continuum at and near the zone center [$\vq_\parallel=(0,0)$] because the interlayer hopping $t_z$ is large.
For larger $t_z$, the continuum in RPA extends to larger energies, and the low-energy plasmons close to the zone center tend to be damped more severely by particle-hole excitations.
This is an important difference with respect to the $t$-$J$-$V$ results where plasmons are always well defined.
Instead, for $q_z=0$ there is no particle-hole continuum near the zone center for any value of $t_z$. 
Comparing Figs. \ref{fig5}(d) and \ref{fig5}(e) with Fig. \ref{fig5}(f), the effect of $t_z$ is less pronounced for the large momentum $\vq=(0.4,0,1)\pi$.
Then, for finite $q_z$ we have to distinguish between two types of regimes.
A finite value of $t_z$ extends the continuum to higher energies, which tends to degrade plasmons at low $\vq_\parallel$, while the continuum from the two-dimensional band structure is the main effect to wash out the plasmon for large $\vq_\parallel$.
Note that the behavior of the particle-hole continuum in Figs. \ref{fig5}(d)-\ref{fig5}(f) is, in general, different to that in Figs. \ref{fig2}(d)-\ref{fig2}(f) for the layered $t$-$J$-$V$ model.
In the latter case the particle-hole continuum is always below the plasmon peak.
In summary, for low-$\vq_\parallel$ momenta such as those discussed in RIXS experiments, the continuum created by a large $t_z$ near the zone center may severely damp the low-energy plasmons in RPA.

As an additional detail of the RPA calculations in Figs. \ref{fig5}(g)-\ref{fig5}(i) we show Re$[1 - V(\vq)\chi^{(0)}(\vq,\omega)]$, i.e., the real part of the denominator of the RPA expression (see Appendix B).
When this denominator crosses zero and $\chi''^{(0)}(\vq,\omega)$ is negligible at the crossing frequency, i.e., no damping, a  plasmon resonance exists.
Note that for the presented values for $\vq_\parallel$ the denominator does not cross zero where the maximum in $\chi''_{RPA}(\vq,\omega)$ occurs (see vertical thin lines in Fig. \ref{fig5}).
This contrasts with the results for the layered $t$-$J$-$V$ model.
In Figs. \ref{fig2}(g)-\ref{fig2}(i) we plot the determinant of the matrix $D^{-1}$ (see Appendix A), which plays the role of the denominator of the RPA for the three selected momenta.
In the layered $t$-$J$-$V$ model the maximum of the peak always matches the frequency where the determinant crosses zero, and no particle-hole continuum exists at this frequency.
See the thin vertical lines in Fig. \ref{fig2}.
Then, in contrast to the $t$-$J$-$V$ results, the features in Figs. \ref{fig5}(a)-\ref{fig5}(c) are not true resonances. 

\section{Discussion and conclusion}

We have demonstrated that if nickelates are cupratelike systems, RIXS experiments are expected to reveal well-defined low-energy plasmons, similar to those recently detected in cuprates and described in the context of the layered $t$-$J$-$V$ model.  
Since the interlayer hopping $t_z$ is likely larger in the infinite-layer nickelates than in the infinite-layer cuprates \cite{botana20}, we expect a larger plasmon gap at the zone center for finite $q_z$ than the gap detected in SLCO \cite{hepting22}.
For instance, while the plasmon gap in SLCO is about $120$ meV, it is expected to be as large as $366$ meV in nickelates. 

In the case of the $t$-$J$-$V$ model why are the low-energy plasmons well defined?
The answer is that the continuum is strongly suppressed by correlations.
This can be seen directly in the equations for the layered $t$-$J$-$V$ model in the Appendix A.
In Eq. (\ref{EktJ}) the strong correlation effects renormalize the hopping parameters by a factor $\delta$, which then reduce the phase space for the particle-hole continuum, allowing the formation of well-defined low-energy plasmons.
A reduction of the bandwidth and the Fermi velocity in comparison with local-density approximation was already discussed in a study of low-energy plasmons in LaCuO$_2$ \cite{falter02}.
In addition, it is important to remark that in the framework of the $t$-$J$ model collective effects are present even if the Coulomb interaction is absent \cite{foussats04}.
That means that spectral weight from the continuum is also removed by this mechanism.
Neither of these two effects occur in a weakly interacting system.

If infinite-layer nickelates are weakly interacting one-band systems or if they are multiorbital with a significant contribution from weakly interacting electrons as expected from Ni orbitals different from the in-plane Ni $d_{x^2-y^2}$, the large interlayer hopping $t_z$ is not renormalized by correlations, and a large particle-hole continuum for finite $q_z$ near the zone center is expected.
Consequently, low-energy plasmons are likely masked.
For example, if the Ni $d_{z^2}$ and the rare-earth orbitals are relevant, they may lead to an enhanced interlayer hopping and make it difficult to yield well-defined plasmons.

In the presence of weakly interacting electrons, as shown in Fig. \ref{fig5}(a)-\ref{fig5}(c), the charge excitation spectrum is broad, and it may be difficult to seprate the plasmon from the broad background.
Notably, previous RIXS studies on infinite-layer nickelates \cite{lu21,tam22,krieger22,rossi22} did not show signs of low-energy plasmons.
However, the experimental conditions in those studies were optimized for the detection of magnetic excitations, i.e., paramagnons, and charge density waves and not for the search of plasmons.
Thus, future RIXS experiments on nickelates are highly desirable and might provide a more conclusive picture about whether low-energy plasmons are present and can be described in nickelates in analogy to cuprates.

It is under debate whether additional orbitals might also play a role at low energies.
Hence, it is not our aim to discuss quantitatively the plasmons, their dispersion, and their possible presence at half filling in multiorbital models.
Our RPA results obtained in the effective single-band model should be regarded at the qualitative level.
If, in the future, more precise experimental data are reported on low-energy plasmons in nickelates, it will be worthwhile to perform calculations in a more realistic model including the multiorbital picture explicitly.

Finally, it will be interesting to find or to develop a layered uncorrelated system with large interlayer hopping.
According to our view, this system should not show well-defined low-energy plasmons, or if they exist, they should be of low intensity and broad and probably difficult to be discerned from the background.
As far as we know, such systems and experiments have not been developed yet; they would provide useful insights about the effect of electron correlations on charge excitations.

\acknowledgments
The authors thank C. Falter, A. M. Ole\'s, A. P. Schnyder, and K.-J. Zhou for fruitful comments.
The authors especially thank to M. Hepting for useful discussions and critical reading of the manuscript.
A.G. thanks the Max-Planck-Institute for Solid State Research in Stuttgart for hospitality and financial support.
H.Y. was supported by JSPS KAKENHI Grants No. JP20H01856.

\appendix

\section{The layered $t$-$J$-$V$ model and the large-$N$ formalism}

The large-$N$ approach for the $t$-$J$ model was originally developed in Ref. \cite{foussats04} and was extensively used in the context of charge excitations in cuprates in, among others, Refs. \cite{bejas12,greco16,greco19,greco20,nag20,hepting22}.
The aim of this section is to give a brief description of the main formulas.

The layer $t$-$J$-$V$ model is written as

\begin{equation}
H = -\sum_{i, j,\sigma} t_{i j}\tilde{c}^\dag_{i\sigma}\tilde{c}_{j\sigma} + 
\sum_{\langle i,j \rangle} J_{ij} \left( \vec{S}_i \cdot \vec{S}_j - \frac{1}{4} n_i n_j \right)
+ \sum_{\langle i,j \rangle} V_{ij} n_i n_j \, 
\label{tJV}  
\end{equation}
where sites $i$ and $j$ run over a three-dimensional lattice. 
The hopping $t_{i j}$ takes a value $t$ ($t'$) between the first (second) nearest-neighbor sites on a square lattice.
The hopping integral between layers is scaled by $t_z$ (see below for the specific form of the electronic dispersion).
$\langle i,j \rangle$ denotes a nearest-neighbor pair of sites.
The exchange interaction $J_{i j}=J$ is considered only inside the plane;
the exchange term between the planes $J_\perp$ is much smaller than $J$ \cite{thio88}.
$V_{ij}$ is the long-range Coulomb interaction on the lattice and is given in momentum space later.
$\tilde{c}^\dag_{i\sigma}$ ($\tilde{c}_{i\sigma}$ ) is the creation (annihilation) operator of electrons with spin $\sigma(=\uparrow, \downarrow)$ in the Fock space without double occupancy.
$n_i=\sum_{\sigma} \tilde{c}^\dag_{i\sigma}\tilde{c}_{i\sigma}$ is the electron density operator, and $\vec{S}_i$ is the spin operator.

In the large-$N$ theory \cite{greco16} the electronic dispersion $\varepsilon_{\vk}$ reads:
\be
\varepsilon_{\vk} = \varepsilon_{\vk}^{\parallel}  + \varepsilon_{\vk}^{\perp} \,,
\label{EktJ}
\ee
where the in-plane dispersion $\varepsilon_{\vk}^{\parallel}$ and the out-of-plane dispersion $\varepsilon_{\vk}^{\perp}$ are given, respectively, by
\begin{eqnarray}
\varepsilon_{\vk}^{\parallel} =& -2 \left( t \frac{\delta}{2}+\Delta \right) (\cos k_{x}+\cos k_{y})
&-4t' \frac{\delta}{2} \cos k_{x} \cos k_{y} - \mu  \,, \label{Epara} \\
\varepsilon_{\vk}^{\perp} =& -2 t_{z} \frac{\delta}{2} (\cos k_x-\cos k_y)^2 \cos k_{z}  \,. \label{Eperp}
\end{eqnarray}

For a given doping $\delta$, the chemical potential $\mu$ and $\Delta$ are determined self-consistently by solving
\begin{equation}{\label {Delta-A}}
\Delta = \frac{J}{4N_s N_z} \sum_{\vk} (\cos k_x + \cos k_y) n_F(\varepsilon_\vk) \; , 
\end{equation}
and 
\begin{equation}
(1-\delta)=\frac{2}{N_s N_z} \sum_{\vk} n_F(\varepsilon_\vk)\,,
\end{equation}
where $n_F$ is the Fermi function, $N_s$ is the total number of lattice sites on the square lattice, and $N_z$ is the number of layers along the $z$ direction.

In the context of the $t$-$J$ model using a path-integral representation \cite{foussats04} for Hubbard operators \cite{hubbard63} a six-component bosonic field is defined as
\begin{equation}
\delta X^{a} = (\delta
R\;,\;\delta{\lambda},\; r^{x},\;r^{y}
,\; A^{x},\;
A^{y})\, ,
\label{boson-field}
\end{equation}
where $\delta R$ describes fluctuations of the number of holes at a given site and thus is related to on-site charge fluctuations, $\delta \lambda$ is the fluctuation of the Lagrange multiplier introduced to enforce the constraint that prohibits the double occupancy at any site, and $r^{x}$ and $r^{y}$ ($A^{x}$ and $A^{y}$) describe fluctuations of the real (imaginary) part of the bond field coming from the $J$-term.

The inverse of the $6\times6$ bare bosonic propagator associated with $\delta X^{a}$ is
\begin{widetext}
\begin{eqnarray}\label{D0}
\left[ D^{(0)}_{ab}({\bf q},\mathrm{i}\omega_{n}) \right]^{-1} = N \left(
 \begin{array}{cccccc}
\frac{\delta^2}{2} \left[ V(\vq)-J(\vq)\right]
& \delta/2 & 0 & 0 & 0 & 0 \\
   \delta/2 & 0 & 0 & 0 & 0 & 0 \\
   0 & 0 & \frac{4}{J}\Delta^{2} & 0 & 0 & 0 \\
   0 & 0 & 0 & \frac{4}{J}\Delta^{2} & 0 & 0 \\
   0 & 0 & 0 & 0 & \frac{4}{J}\Delta^{2} & 0 \\
   0 & 0 & 0 & 0 & 0 & \frac{4}{J}\Delta^{2} \
 \end{array}
\right),
\end{eqnarray}
\end{widetext}

\noindent where  $\vq$ is a three dimensional wavevector and $\omega_n$ is a bosonic Matsubara frequency. 
$J(\vq) = \frac{J}{2} (\cos q_x +  \cos q_y)$, and $V(\vq)$ is the long-range Coulomb interaction for a layered 
system,

\be
V(\vq)=\frac{V_c}{A(q_x,q_y) - \cos q_z} \,,
\label{LRC}
\ee
where $V_c= e^2 d(2 \epsilon_{\perp} a^2)^{-1}$ and 
\be
A(q_x,q_y)=\alpha (2 - \cos q_x - \cos q_y)+1 \,.
\ee
These expressions are easily obtained by solving Poisson's equation on the lattice \cite{becca96}.
Here $\tilde{\epsilon}=\epsilon_\parallel/\epsilon_\perp$, and $\epsilon_\parallel$ and $\epsilon_\perp$ are the dielectric constants parallel and perpendicular to the planes, respectively.
$e$ is the electric charge of electrons; $\alpha=\frac{\tilde{\epsilon}}{(a/d)^2}$, and $a$ is the lattice spacing in the planes.
The in-plane momentum $\qp=(q_x,q_y)$ is measured in units of $a^{-1}$.
Similarly, $d$ is the distance between the planes, and the out-of-plane momentum $q_z$ is measured in units of $d^{-1}$.
In the present work we consider $V_c$ and $\alpha$ to be independent parameters.

At leading order, the bare propagator $D^{(0)}_{ab}$ is renormalized in $O(1/N)$.
From the Dyson equation the renormalized bosonic propagator is
\be
[D_{ab}(\vq,\mathrm{i}\omega_n)]^{-1}
= [D^{(0)}_{ab}(\vq,\mathrm{i}\omega_n)]^{-1} - \Pi_{ab}(\vq,\mathrm{i}\omega_n)\,.
\label{dyson}
\ee

Here the $6 \times 6$ boson self-energy matrix $\Pi_{ab}$ is
\begin{widetext}
\begin{eqnarray}
&& \Pi_{ab}(\vq,\mathrm{i}\omega_n)
            = -\frac{N}{N_s N_z}\sum_{\vk} h_a(\vk,\vq,\varepsilon_\vk-\varepsilon_{\vk-\vq}) 
            \frac{n_F(\varepsilon_{\vk-\vq})-n_F(\varepsilon_\vk)}
                                  {\mathrm{i}\omega_n-\varepsilon_\vk+\varepsilon_{\vk-\vq}} 
            h_b(\vk,\vq,\varepsilon_\vk-\varepsilon_{\vk-\vq}) \nonumber \\
&& \hspace{25mm} - \delta_{a\,1} \delta_{b\,1} \frac{N}{N_s N_z}
                                       \sum_\vk \frac{\varepsilon_\vk-\varepsilon_{\vk-\vq}}{2}n_F(\varepsilon_\vk) \; ,
                                       \label{Pi}
\end{eqnarray}
\end{widetext}
where the six-component interaction vertex is given by
\begin{widetext}
\begin{align}
 h_a(\vk,\vq,\nu) =& \left\{
                   \frac{2\varepsilon_{\vk-\vq}+\nu+2\mu}{2}+
                   2\Delta \left[ \cos\left(k_x-\frac{q_x}{2}\right)\cos\left(\frac{q_x}{2}\right) +
                                  \cos\left(k_y-\frac{q_y}{2}\right)\cos\left(\frac{q_y}{2}\right) \right];1;
                 \right. \nonumber \\
               & \left. -2\Delta \cos\left(k_x-\frac{q_x}{2}\right); -2\Delta \cos\left(k_y-\frac{q_y}{2}\right);
                         2\Delta \sin\left(k_x-\frac{q_x}{2}\right);  2\Delta \sin\left(k_y-\frac{q_y}{2}\right)
                 \right\} \, .
\label{vertex-h}
\end{align}
\end{widetext}

As discussed previously \cite{foussats04,bejas12}, the element $(1,1)$ of $D_{ab}$ is related to the usual charge-charge correlation function $\chi_{c} (\vr_i -\vr_j, \tau)=\bra T_\tau n_i(\tau) n_j(0)\ket$, which in the large-$N$ scheme is computed in the $\vq$-$\omega$ space as 
\begin{eqnarray}\label{CH}
\chi_{c}(\vq,\mathrm{i}\omega_n)= N \left ( \frac{\delta}{2} \right )^{2} D_{11}(\vq,\mathrm{i}\omega_n)  \,.
\end{eqnarray}
\noindent$\chi_{c}(\vq,\mathrm{i}\omega_n)$, i.e., $D_{11}(\vq,\mathrm{i}\omega_n)$ contains the plasmons.
In the large-$N$ approach the particle-hole continuum is given by $\Pi_{22}(\vq,\mathrm{i}\omega_n)$ [see Eq. (\ref{Pi})].
It is important to remark that the charge-charge correlation function is nearly unaffected by the value of $J$ \cite{bejas17}.

After performing the analytical continuation $\mathrm{i}\omega_n \rightarrow \omega+\mathrm{i} \Gamma$ in \eq{CH}, we obtain the imaginary part of the charge-charge correlation function $\chi''_{c}(\vq,\omega)$, which can be directly compared with RIXS.
While $\Gamma$ ($>0$) is infinitesimally small, we employ a finite broadening $\Gamma$ \cite{greco19,greco20}.
This $\Gamma$ mimics not only effects of the experimental resolution but also a broadening of the spectrum due to electron correlations
\cite{prelovsek99}, so that we can successfully reproduce the peak width of experimental data \cite{greco19,nag20,hepting22}.

\section{Random phase approximation}

In the RPA the charge correlation function is given by the well-known expression \cite{mahan}
\begin{eqnarray}\label{chiRPA}
\chi_{RPA}({\bf q},\mathrm{i}\omega_{n})=\frac{\chi^{(0)}({\bf q},\mathrm{i}\omega_{n})}
{1-V(\vq)\chi^{(0)}({\bf q},\mathrm{i}\omega_{n})} \, , 
\end{eqnarray}
\noindent  where $\chi^{(0)}({\bf q},\mathrm{i}\omega_{n})$ is the usual Lindhard function, which accounts for the particle-hole continuum.
The electron dispersion $E_{\vk}$ is
\be
E_{\vk} = E_{\vk}^{\parallel}  + E_{\vk}^{\perp} \,,
\label{Ek}
\ee
where 
\begin{eqnarray}
E_{\vk}^{\parallel} =-2 t(\cos k_{x}+\cos k_{y})
-4t'\cos k_{x} \cos k_{y} - \mu  \,, \label{EparaRPA} \\
E_{\vk}^{\perp} = -2 t_{z} (\cos k_x-\cos k_y)^2 \cos k_{z}  \,. \label{EperpRPA}
\end{eqnarray}

While in RPA the bare electronic dispersion (\ref{Ek}) must be used, in the $t$-$J$ model [Eq. (\ref{EktJ})] the presence of the doping 
$\delta$ and $\Delta \sim J$ shows the presence of correlations effects in the electron dispersion. 

We have considered only the long-range Coulomb interaction $V(\vq)$ in Eq.~(\ref{chiRPA}).
One might wish to include other interactions such as a local Coulomb interaction $U$ and a nearest-neighbor density-density interaction.
The inclusion of these interactions, as long as the system remains in the paramagnetic homogeneous phase as in our case, does not change the main conclusions; it may yield a small change in the value of $V_c$.

\bibliography{main} 

\end{document}